\documentclass[aip,amsmath,amssymb,reprint]{revtex4-1}
\usepackage{xcolor}
\usepackage{graphicx}
\usepackage{dcolumn}
\usepackage{bm}

\usepackage[utf8]{inputenc}
\usepackage[T1]{fontenc}
\usepackage{mathptmx}
\usepackage{etoolbox}
\usepackage{adjustbox}
\usepackage{multirow}
\makeatletter
\begin{document}

\title{Methane hydrate nucleation frustration and dimensional reduction of structural order under nanoconfinement}

\author{José Torres-Arenas}
\affiliation{División de Ciencias e Ingenierías, Universidad de Guanajuato}
\affiliation{Departamento de Física Aplicada, Universidade de Vigo, E-36310 Vigo Spain.}

\author{Ángel M. Fernández-Fernández }
\affiliation{Departamento de Física Aplicada, Universidade de Vigo, E-36310 Vigo Spain.}

\author{Martín Pérez-Rodríguez}
\affiliation{Instituto de Química Física Blas Cabrera, CSIC, E-28006 Madrid, Spain.}

\author{Manuel M. Piñeiro}
\affiliation{Departamento de Física Aplicada, Universidade de Vigo, E-36310 Vigo Spain.}

\date{\today}

\begin{abstract}

Methane hydrate nucleation under nanoconfinement remains poorly understood due to the complex interplay between geometric restriction and molecular ordering. Here, we investigate the structural organization of water–methane systems confined between silica planar slit pores with widths ranging from 1 to 5 nm and temperatures between 250 and 295 K. Three-dimensional radial distribution functions reveal a clear suppression of hydrate-like ordering at strong confinement ($<$ 2 nm), indicating frustrated nucleation. In contrast, projected two-dimensional correlations exhibit pronounced in-plane structural organization, evidencing a confinement-induced reduction in the dimensionality of molecular order.
\end{abstract}
\maketitle

Methane hydrates are crystalline inclusion compounds formed by hydrogen-bonded water cages encapsulating methane, typically under low-temperature and high-pressure conditions\cite{Sloan2008}. 
Such conditions naturally occur in marine sediments along continental margins and in permafrost regions, where methane hydrates constitute a vast carbon reservoir\cite{Max2011,Chong20161633}. 
In seabed environments, hydrate formation is strongly affected by confinement within sediment pores, mineral surfaces, and interfacial effects, which can substantially modify nucleation pathways and stability compared to bulk systems. Understanding the role of nanoconfinement is therefore essential 
for evaluating methane hydrates in the context of climate, the global carbon cycle, energy recovery, 
and seafloor stability\cite{Zhang2025}.
Despite extensive experimental and computational efforts\cite{Bagherzadeh20123188,Zheng2022124718,Zhang2022, Zhang2023,Fan2024,Li2026}, the microscopic mechanisms governing methane hydrate nucleation under confinement remain poorly understood, particularly in terms of how spatial restriction modifies the development of hydrate-like structural order. In this work, we investigate water–methane systems confined between planar walls with pore sizes ranging from 1 to 5 nm over a broad temperature range (250–295 K). By combining three-dimensional and projected radial distribution functions, we show that strong confinement suppresses bulk-like hydrate ordering while promoting pronounced in-plane structuring. Our results reveal that nanoconfinement induces a reduction in the dimensionality of molecular order, providing a structural origin for the frustration of methane hydrate nucleation.\\
Here, we demonstrate using atomistic Molecular Dynamics (MD) simulations that the frustration of methane hydrate nucleation under nanoconfinement originates from a confinement-induced mismatch between local structural ordering and the three-dimensional connectivity required for hydrate formation. By systematically varying pore size and temperature, we identify a regime of strong confinement in which bulk-like hydrate signatures vanish in three-dimensional correlations while in-plane ordering persists. This dimensional decoupling provides a clear structural framework to rationalize nucleation inhibition in confined environments and highlights the fundamental role of spatial constraints in controlling phase transitions at the nanoscale.\\

In previous works\cite{Fernandez2022JML, Fernandez2024JCP}, 
we have analyzed using MD the crystallization and growth of methane hydrate starting from an initial seed confined within an atomistic silica slit pore of $\approx$ 7 nm width. The results evidenced complete crystallization, showing structural defects and disclinations, as well as the formation of a remarkable interfacial fluid water layer adsorbed at the pore walls. 

These findings naturally raise two important questions: whether methane hydrate can nucleate spontaneously under such confined conditions, and  whether a limiting pore size exists below which hydrate nucleation becomes frustrated. \\

The spontaneous nucleation of methane hydrate under bulk conditions has been studied using MD simulations by different authors,\cite{Walsh20091095,Sarupria20122942,Khurana201711176,Chen2021} who have highlighted the key role of local structural ordering, density fluctuations, and cage-like precursor formation in driving the nucleation process.

In this work, we investigate methane hydrate formation under nanoconfinement by considering a system composed of water and methane molecules confined between two planar crystalline walls of $\alpha$-silica. 
Water molecules are modeled using the widely adopted TIP4P/Ice force field,\cite{Abascal2005Ice} 
which accurately reproduces the phase behavior of ice and hydrate-forming systems. Methane molecules are described using the All-Atom Optimized Potentials for Liquid Simulations (OPLS-AA) force field.\cite{MacKerellJr.19983586} The confinement is imposed along the $z$-direction by varying the distance between the two silica walls, thereby defining slit pores with widths ranging from 1 to 5 nm. Each pore is filled with water and methane molecules while maintaining a constant composition across all systems. A more detailed description of the system setup, including surface structure and simulation protocols, can be found in our previous works.\cite{Fernandez2022JML, Fernandez2024JCP}

In this context, we investigate how nanoconfinement modifies molecular ordering by combining three-dimensional and projected radial distribution functions with local bond order parameters, thereby identifying the structural origin of methane hydrate nucleation frustration.

\begin{figure}[t]
\includegraphics[width=\columnwidth]{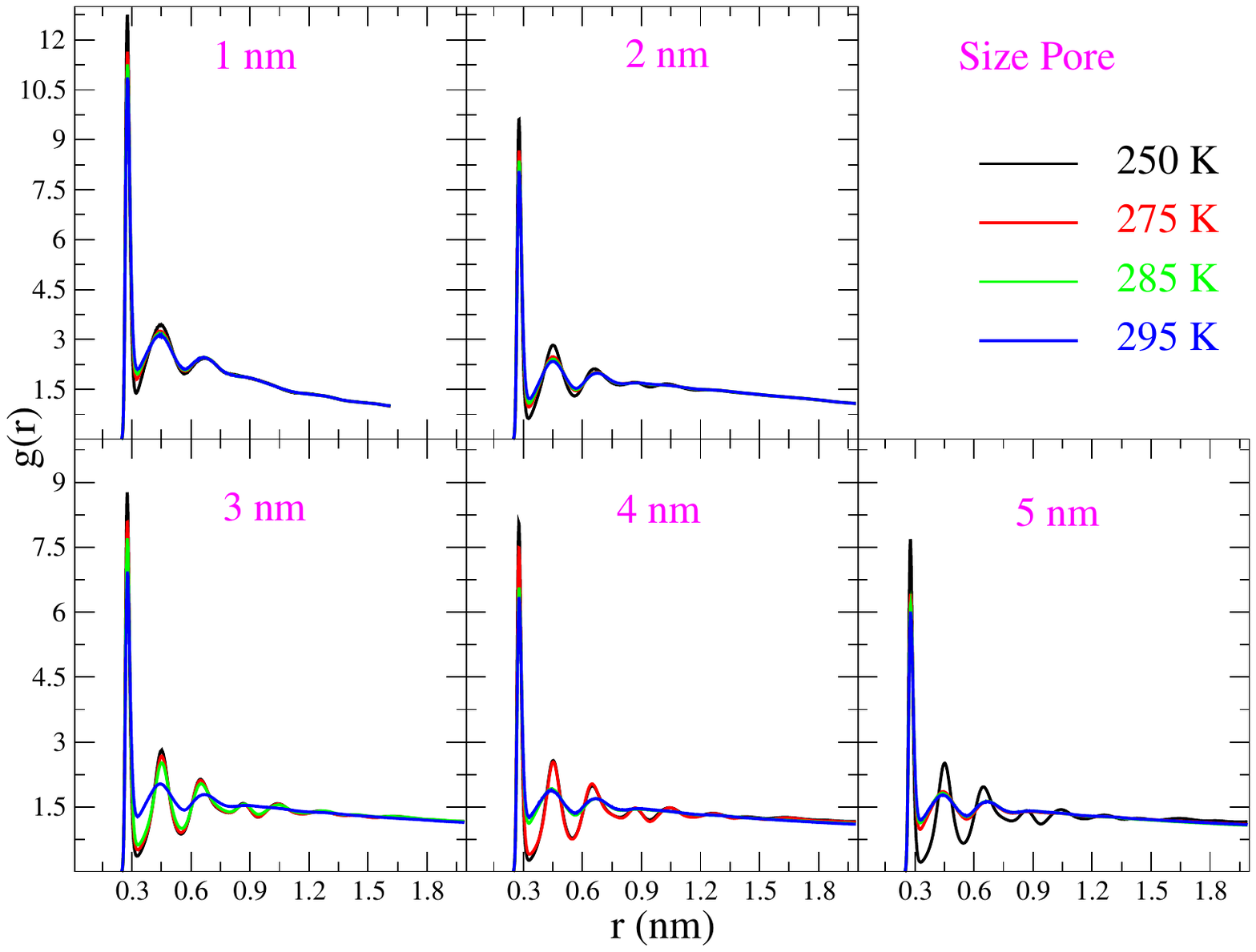}
\caption{Three-dimensional radial distribution functions, $g(r)$, for methane–water systems confined in planar silica pores with widths from 1 to 5 nm at temperatures between 250 and 295 K. Top panels: 1 nm (left) and 2 nm (right). Bottom panels: 3 nm (left), 4 nm (center), and 5 nm (right). Strong confinement (1–2 nm) suppresses three-dimensional structural ordering at all temperatures, whereas larger pores (3–5 nm) exhibit hydrate-like features at low temperatures that progressively vanish upon heating.}
\label{rdf3d}
\end{figure}

The three-dimensional radial distribution functions (RDFs) provide a first structural characterization of methane–water ordering under nanoconfinement 
across pore sizes (1--5 nm) and temperatures (250--295 K), see Figure \ref{rdf3d}. 
Despite the limitations associated with spherical averaging in anisotropic confined systems, clear trends in structural organization can be identified. For the strongest confinement cases (1 and 2 nm), the RDFs show a complete suppression of hydrate-like structural signatures over the entire temperature range studied. The absence of well-defined coordination shells is consistent with the loss of three-dimensional ordering, as further confirmed by the structural order parameters 
F$_3$\cite{BAEZ1994177} and F$_4$\cite{Rodger1996326} (discussed below). In contrast, the 3 nm pore exhibits pronounced structural features between 250 and 285 K, which disappear at 295 K, 
revealing a strong thermal sensitivity of confined ordering. A similar trend is observed for the 4 nm pore, where hydrate-like structuring persists only at 250 and 275 K. 
For the 5 nm pore, structural ordering is retained solely at 250 K, indicating a progressive destabilization of hydrate-like organization with increasing temperature. Overall, these results demonstrate a combined effect of confinement and temperature on the stability 
of three-dimensional hydrate-like ordering.\\

\begin{figure}[t]
\includegraphics[width=\columnwidth]{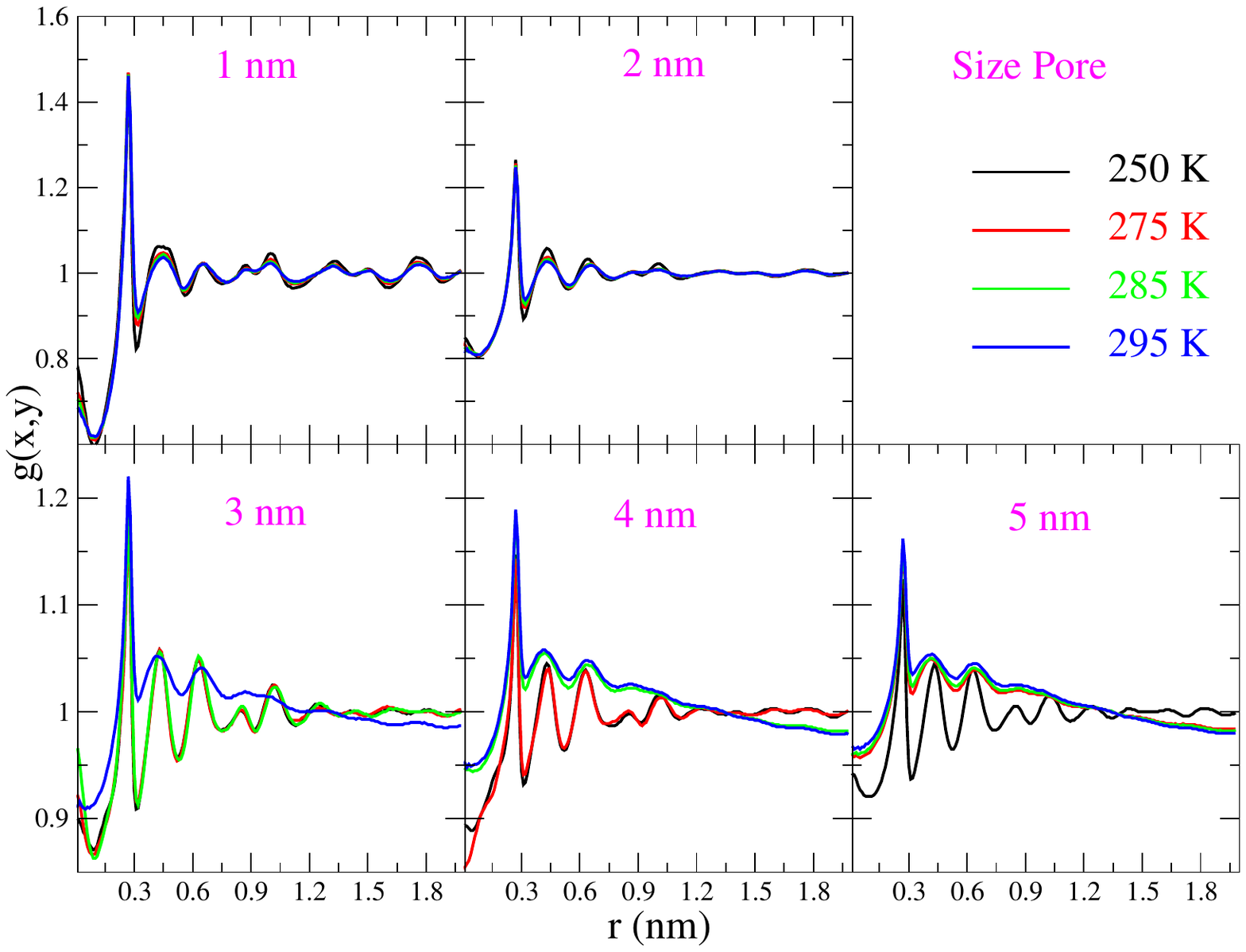}
\caption{Projected radial distribution functions, $g(x,y)$, for methane–water systems confined in planar silica pores with widths from 1 to 5 nm at temperatures between 250 and 295 K. Top panels: 1 nm (left) and 2 nm (right). Bottom panels: 3 nm (left), 4 nm (center), and 5 nm (right). For pore sizes of 3–5 nm, the projected correlations follow the same trends as the three-dimensional RDFs. 
In contrast, under strong confinement (1–2 nm), pronounced in-plane structural ordering persists 
at all temperatures, despite the absence of three-dimensional hydrate-like order.}
\label{rdf2D}
\end{figure}

To further elucidate the nature of structural ordering under confinement, we analyze the projected radial distribution functions in the lateral (x–y) plane, which provide insight into in-plane molecular organization along the confining surfaces, provided in Figure \ref{rdf2D}.
For pore widths between 3 and 5 nm, the projected g(x,y) functions closely mirror the behavior observed in the three-dimensional RDFs, confirming that the detected structural features correspond to genuinely three-dimensional ordering. In these cases, hydrate-like signatures persist only under the same combinations of low temperature and intermediate confinement identified previously, with no evidence of additional ordering emerging from the projection.
In striking contrast, a fundamentally different behavior emerges under strong confinement (1 and 2 nm pores). While the three-dimensional RDFs indicate a complete loss of hydrate-like ordering in these systems, the projected g(x,y) functions reveal the presence of pronounced and well-defined in-plane structural ordering across all investigated temperatures. This persistent lateral ordering indicates that, even under conditions where three-dimensional connectivity is suppressed, the system retains a high degree of local structural organization parallel to the confining walls.
This clear decoupling between in-plane and three-dimensional ordering demonstrates that strong nanoconfinement does not eliminate molecular structure, but rather restructures it into a two-dimensional arrangement incompatible with the formation of a three-dimensional hydrate network.\\

\begin{figure}[t]
\includegraphics[width=\columnwidth]{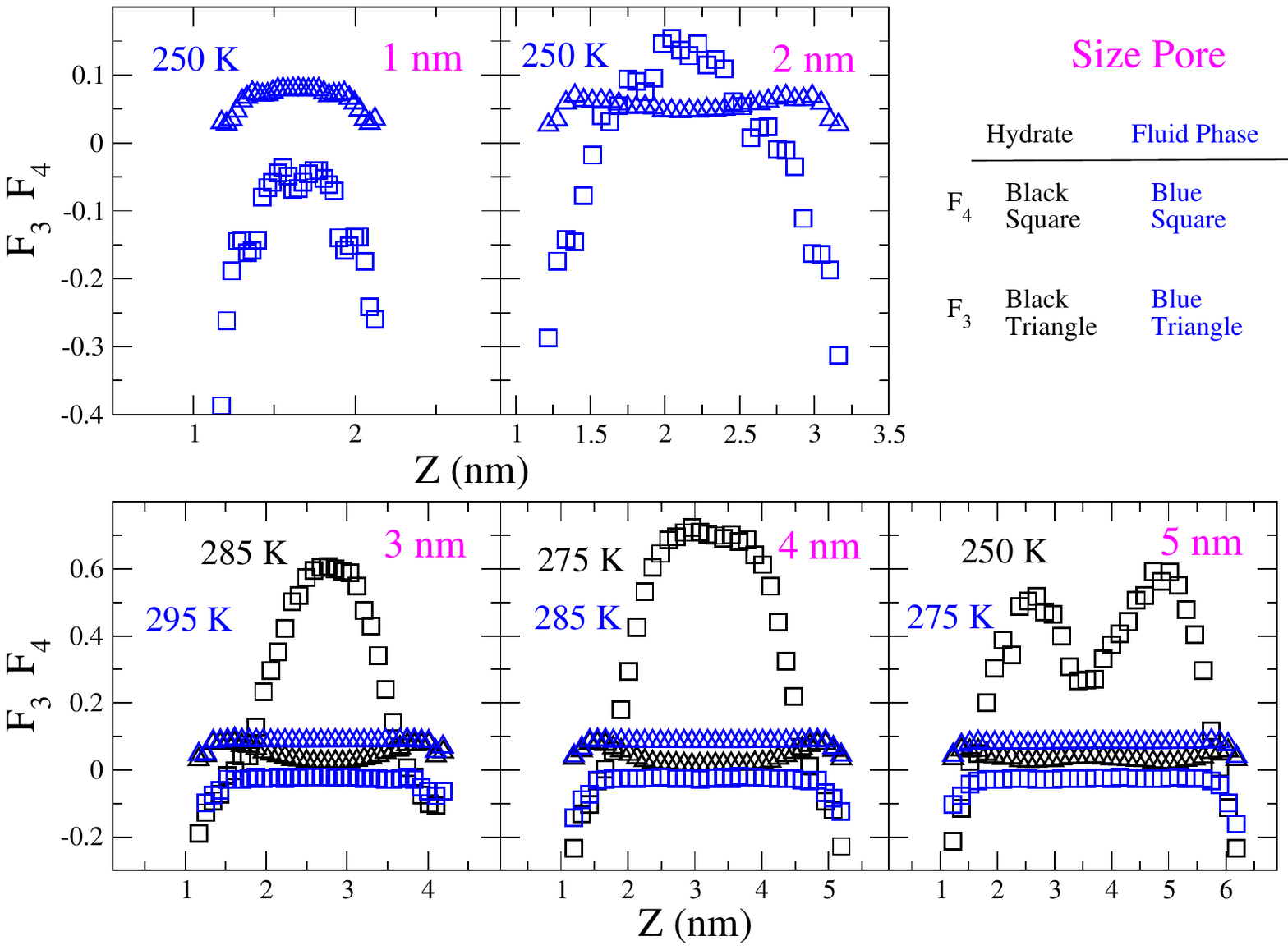}
\caption{Spatial profiles of the structural order parameters F$_3$ and F$_4$ for methane--water systems confined in planar silica pores of widths between 1 and 5 nm. Top panels: 1 nm (left) and 2 nm (right) pores at $T=250$ K. Bottom panels: 3 nm (left), 4 nm (center), and 5 nm (right) pores evaluated at the pairs of consecutive temperatures where the transition 
between structured and unstructured regimes is observed. Strong confinement (1--2 nm) suppresses hydrate-like ordering, as indicated by the nearly uniform behavior of F$_3$ 
and the predominantly negative values of F$_4$. For larger pores (3--5 nm), both parameters capture the temperature-driven loss of structural ordering, consistent with the behavior observed in the radial distribution functions.}
\label{order}
\end{figure}

To quantitatively characterize the emergence and suppression of hydrate-like ordering under confinement, we compute the structural order parameters F$_3$ and F$_4$ for all pore sizes and relevant thermodynamic conditions. These are given in Figure \ref{order}. The results are presented in a composite figure, where strongly confined systems (1 and 2 nm pores) are analyzed only at $T=250$ K, since the absence of hydrate-like ordering under the most favorable thermodynamic conditions implies its suppression at higher temperatures as well. 
In contrast, intermediate confinement cases (3--5 nm pores) exhibit clear temperature-dependent structural transitions in the RDFs. Accordingly, F$_3$ and F$_4$ are evaluated for pairs of consecutive temperatures corresponding to the onset and disappearance of hydrate-like ordering for each pore size.

For the most strongly confined systems (1 and 2 nm), the analysis at 250 K confirms the suppression of hydrate-like ordering. For the 1 nm pore, the F$_3$ parameter remains nearly constant throughout the pore interior, with only a slight decrease near the confining walls. Consistently, F$_4$ remains negative across the entire pore, indicating the absence of tetrahedral hydrate-like organization. A similar behavior is observed for the 2 nm pore, where F$_3$ presents a slightly lower central value, followed by a weak increase toward intermediate regions of the pore and a subsequent decrease close to the walls. In contrast to the 1 nm case, F$_4$ exhibits a profile with a weak positive maximum of approximately 0.15 at the pore center, progressively decreasing toward negative values near the walls. Despite this slight order enhancement in the central region, the magnitude and spatial distribution of both parameters remain incompatible with the formation of stable three-dimensional hydrate structures under strong confinement.

For larger pore sizes, the order parameters clearly capture the transition between structured and unstructured regimes as a function of temperature. For the 3 nm pore, hydrate-like ordering persists up to 285 K, where both F$_3$ and F$_4$ still exhibit noticeable spatial modulation across the pore, indicating the presence of organized three-dimensional structure. However, at 295 K both parameters become significantly flatter, revealing the collapse of structural ordering, in agreement with the corresponding RDFs. A similar behavior is observed for the 4 nm pore, where the transition occurs between 275 K and 285 K. At 275 K, both order parameters retain clear spatial variation, while at 285 K this modulation is strongly reduced, indicating the disappearance of hydrate-like organization. For the 5 nm pore, the structural transition shifts further toward lower temperatures, taking place between 250 K and 275 K. In this case, hydrate-like ordering is only preserved at the lowest temperature, whereas both F$_3$ and F$_4$ lose their characteristic spatial structure upon heating. Overall, the progressive shift of the transition temperature toward lower values with increasing pore size demonstrates that confinement strongly modifies the stability window of methane hydrate ordering. The behavior observed for F$_3$ and F$_4$ is fully consistent with the trends identified in both the three-dimensional and projected radial distribution functions, with strong confinement fully suppressing hydrate-like organization and larger pores exhibiting progressively temperature-driven structural destabilization.\\

The combined analysis of three-dimensional and projected radial distribution functions, together with the spatially resolved order parameters F$_3$ and F$_4$, provides a consistent molecular-level picture of methane hydrate formation under nanoconfinement. Strong confinement ($\leq$2 nm) leads to a complete suppression of three-dimensional hydrate-like ordering, as evidenced by the absence of structural signatures in g(r) and the nearly uniform behavior of the order parameters across the pore. However, the projected g(x,y) functions reveal that this apparent structural loss is accompanied by the persistence of pronounced in-plane ordering, indicating that molecular organization is not eliminated but instead restricted to two dimensions. In contrast, for larger pore sizes (3–5 nm), hydrate-like structural features emerge in three dimensions within a temperature-dependent stability window, which progressively narrows with decreasing confinement. This crossover between structured and unstructured regimes is consistently captured by both RDFs and order parameters, and shifts systematically with pore size and temperature. Overall, these results demonstrate that nanoconfinement does not simply disrupt ordering, but fundamentally alters its dimensionality, leading to a decoupling between local in-plane structuring and the three-dimensional connectivity required for hydrate nucleation. This dimensional reduction provides a clear structural mechanism for the frustration of methane hydrate formation under confinement.

These results provide a coherent molecular-level picture of how confinement controls methane hydrate formation. Our analysis shows that methane hydrate nucleation can occur spontaneously in planar silica pores, with surface-induced ordering promoting structural organization under moderate confinement. However, this effect is non-monotonic, and a critical pore width of approximately 2 nm 
is identified below which hydrate-like ordering is completely suppressed, leading to nucleation frustration. Under this strong confinement regime, the system does not become fully disordered. Instead, pronounced water layering and persistent in-plane ordering emerge, revealing a confinement-induced reduction in the dimensionality of molecular organization, where three-dimensional hydrate connectivity is replaced by two-dimensional structural ordering. For larger pores ($\geq 3$ nm), hydrate-like structures are recovered within a temperature-dependent stability window 
that progressively narrows with increasing confinement. Overall, these findings demonstrate that nanoconfinement governs methane hydrate formation through a competition between interfacial ordering and suppression of three-dimensional connectivity, providing a clear structural origin for hydrate nucleation frustration.

\section*{Author Contributions}
All authors contributed equally to this work. All authors have read and approved the final version of the manuscript.

\section*{Conflicts of interest}
There are no conflicts to declare.

\section*{Data availability}
The data supporting the findings of this study are available from the corresponding author upon reasonable request.

\section*{Acknowledgements}

 MMP and AMFF acknowledge grant Refs.~PID2021-125081NB-I00 and PID2024-158030NB-I00 funded by MCIN/AEI/10.13039/501100011033 (Spain), FEDER EU and Programa de axudas de apoio á etapa de formación posdoutoral da Xunta de Galicia (Consellería de Cultura, Educación, Formación Profesional e Universidades) (orde 27/12/2023) P.P. 0000 421S 140.08. We also  acknowledge computing resources provided by CESGA (Santiago de Compostela, Spain). 
JTA would like to thank the Departamento de F\'isica Aplicada of the Universidade de Vigo and CESGA for all the facilities provided  to a sabbatical stay, during which, this work was developed. Also to financial support of Universidad de Guanajuato and SECIHTI (through SNII), that makes this stay possible.

\bibliography{biblio}

\end{document}